\documentclass{article}

\usepackage[nonatbib, final]{neurips_2020}
\usepackage[utf8]{inputenc} % allow utf-8 input
\usepackage[T1]{fontenc}    % use 8-bit T1 fonts
\usepackage{hyperref}       % h\textbf{}yperlinks
\usepackage{url}            % simple URL typesetting
\usepackage[table]{xcolor}
\usepackage{booktabs}       % professional-quality tables
\usepackage{amsfonts}       % blackboard math symbols
\usepackage{nicefrac}       % compact symbols for 1/2, etc.
\usepackage{microtype}      % microtypography
\usepackage[sorting=none,maxbibnames=99]{biblatex}
\usepackage{pgfplotstable}
\usepackage{tikz}
\usepackage{amsmath,amssymb}

\usepackage{wrapfig}
\usepackage{subfig}
\usepackage{float}
\usepackage{changepage}
\usepackage{todonotes}
\usepackage{enumitem}
\DeclareUnicodeCharacter{0301}{}

\usepackage{physics}
\usepackage{amsmath}
\usepackage{tikz}
\usepackage{mathdots}
\usepackage{yhmath}
\usepackage{cancel}
\usepackage{color}
\usepackage{siunitx}
\usepackage{array}
\usepackage{multirow}
\usepackage{amssymb}
\usepackage{gensymb}
\usepackage{tabularx}
\usepackage{booktabs}
\usetikzlibrary{fadings}
\usetikzlibrary{patterns}
\usetikzlibrary{shadows.blur}
\usetikzlibrary{shapes}

\usepackage{amsmath,amsfonts,bm}

\title{Structure-aware generation of drug-like molecules}
\author{%
    Pavol Drotár\\
    University of Cambridge\\
    \texttt{pd451@cantab.ac.uk}\\
    \And
    Arian Rokkum Jamasb, Ben Day, Cătălina Cangea, Pietro Liò \\
    University of Cambridge\\
    \texttt{\{arj39,bjd39,ccc53,pl219\}@cam.ac.uk}\\
}

\begin{document}
\maketitle
\begin{abstract}
\vspace*{-0.2cm}
Structure-based drug design involves finding ligand molecules that exhibit structural and chemical complementarity to protein pockets. Deep generative methods have shown promise in proposing novel molecules from scratch (\emph{de-novo} design), avoiding exhaustive \emph{virtual screening} of chemical space. Most generative de-novo models fail to incorporate detailed ligand-protein interactions and 3D pocket structures. We propose a novel supervised model that generates molecular graphs jointly with 3D pose in a discretised molecular space. Molecules are built atom-by-atom inside pockets, guided by structural information from crystallographic data. We evaluate our model using a docking benchmark and find that guided generation improves predicted binding affinities by $8\%$ and drug-likeness scores by $10\%$ over the baseline. Furthermore, our model proposes molecules with binding scores exceeding some known ligands, which could be useful in future wet-lab studies.
\end{abstract}
\vspace*{-0.2cm}
\label{section:introduction}
\section{Introduction}
\vspace*{-0.2cm}
Increasing attrition rates in drug design \cite{gaudelet2021utilizing, drug_costs, developability} will require higher automation \cite{vs_challenges} for screening the vast space of $\sim10^{60}$ drug-like chemicals \cite{chemical_space} and proposing novel substances. Modern \textit{de-novo} design methods build ligands incrementally from small fragments and use scoring functions to predict their in-pocket binding affinities \cite{de_novo}. Early generative methods designed SMILES \cite{smiles}, string-based molecular representations, using language models, RNNs, and GANs \cite{gvae, molecular_autoencoder, reinvent, neil2018exploring, molgan}, discarding molecular graph information. Recently, graph neural networks have shown sufficient expressivity to capture meaningful molecular graph features \cite{cgvae, molecular_autoencoder, deep_generative_models, mpnn}. Most generative models aim to produce molecules resembling datasets such as ZINC \cite{zinc} (\textit{ligand-based} design) and are evaluated based on similarity metrics \cite{virtual_screening}. These objectives do not capture the real difficulty of designing novel drug-like molecules that would bind to a target protein \cite{dock}.

Some approaches condition generation on vector descriptors of pockets. DeLinker \cite{delinker} designs fragment linkers (carbon-based small molecules) based on scalar distances. DEVELOP \cite{develop}, SyntaLinker \cite{syntaLinker}, and \textcite{protein_shape_gan, ligand_shape_gan} incorporate 3D CNN pocket encodings. However, pocket encodings are only used at initialisation and no model incorporates high-resolution 3D pocket structures available in crystallographic datasets \cite{dude, pdb_bind}, which could be used to generate molecules directly in-pocket. We draw on approaches that generate 3D molecular conformers and we explicitly add pocket-awareness. \textcite{simm_conformers} introduce rotation-and-translation invariant coordinates to deep conformer generation and MolGym \cite{molgym} successfully models atomic interactions in 3D using SchNet \cite{schnet}.  The direct generation of molecules in 3D as point-clouds comes with a larger unconstrained search space and we use an underlying graph to incorporate valency constraints as in CGVAE \cite{cgvae}.

Our model generates the coordinates and graph jointly --- molecules are generated atom-by-atom (similar to \cite{cgvae, deep_generative_models}) and 3D atom positions are predicted in discretised internal coordinates. Interactions with the pocket are modeled using geometric deep learning (SchNet \cite{schnet}) and updated at each generative step to supervise generation. We optimise the generated conformers using RDKit \cite{distance_geometry} and evaluate the model using a docking benchmark proposed to capture the full difficulty of ligand design \cite{dock}. To our knowledge, this is the first supervised method for joint molecular graph and pose generation.

%\begin{figure}[h]
%\centering
%\includegraphics[scale=0.12,trim={1cm 0 0 1cm},clip]{figures/abstract.pdf}
%\caption{A potential drug to test proposed by our model, docked inside the %target protein.}
%\label{fig:abstract}
%\end{figure}
\label{section:methods}
\section{Methods}
\vspace*{-0.2cm}
\label{subsection:representation}
\paragraph{Ligand and protein representation.}
We use the PDBBind crystallographic dataset \cite{pdb_bind} which contains co-crystallised structures of proteins and ligands. We represent molecules as graphs $G(V,E)$ over atoms and bonds where vertex features are one-hot atom types $\bm{l_i} \in \mathbb{L} = \{\textup{C, F, N, Cl, O, I, P, Br, S, H}\}$ and edge features are bond types $\bm{e}_{ij} \in \mathbb{B} = \{\textup{single, double, triple, aromatic}\}$. We also have 3D Euclidean coordinates $\bm{r_i} \in \mathbb{R}^3$ for each atom, however, we use internal molecular coordinates ($z$-matrix) during molecule generation. These are invariant under translations and rotations of a molecule. Each atom is described in terms of the bond length $d$, bond angle $\alpha$, and dihedral (planar) angle $\theta$ with respect to the three nearest atoms (Figure \ref{fig:torsion}, Appendix \ref{app:appendix_coords}). We discretise the bond and dihedral angles according to PDB dataset statistics to reduce coordinate space and, as we observe in evaluation, this makes coordinate prediction feasible.

\label{subsection:graph_autoencoders}
\paragraph{Sequential variational graph auto-encoding.}
Variational graph autoencoders are trained to encode graphs into a latent embedding from which the graph can be reconstructed. The aim is to then use the decoder to generate new graphs by sampling the latent space. The encoder embeds a graph $G$ into a latent representation $\mathbf{Z} \sim q_{\phi}(\mathbf{Z}|G)$ parameterised as a normal distribution. The decoding process optimises a distribution $p_{\theta}(G|\mathbf{Z})$ to re-generate the original graph by conditioning on the latent embedding. We decompose the distribution over graphs using atom-by-atom (sequential) generation, producing graphs $G^0, G^1, ..., G^T$, each conditioned on the previous graph and the latent representation $G^{t+1} \sim p(G^{t+1}|G^t,\mathbf{Z})$. Sequential generation has been shown to be effective in generating molecules \cite{deep_generative_models, gran} and allows granular valency constraints to be encoded \cite{cgvae}. The model is trained using an Evidence Lower-Bound (ELBO) objective \cite{variational_bayes} on a set of training graphs.

\label{subsection:teacher_forcing}
\paragraph{Teacher forcing.}
In decoding, partial graphs are constructed by adding a new node, edge, or a coordinate. Existing approaches condition each partial graph on the full generation history $G^{t+1} \sim p(G^{t+1}|G^t, \dots, G^0, \mathbf{Z})$, however, this is unnecessary for molecules as there is no additional information in the generation history, and conditioning solely on the latest partial graph $G^t$ scales to larger molecules \cite{cgvae, deep_generative_models}. It is impractical to supervise the process in an end-to-end fashion due to discrete node and edge selection construction steps. Instead, we use teacher forcing and supervise each step \cite{cgvae, delinker} using a predefined graph construction order during training. We use a breadth-first order of node additions that has been demonstrated to work well for molecules \cite{cgvae}.

\label{section:approach}
\section{Structure-aware generative model}
\vspace*{-0.2cm}
The model can be started in \textit{generation} mode, which leverages a trained decoder to construct a new molecule from a randomly sampled latent encoding, or in \textit{training} mode, where the encoder-decoder setup is supervised using ground truth molecule construction steps.

\begin{figure}[htp]
\centering
\includegraphics[trim = 10 15 45 15, width = \textwidth]{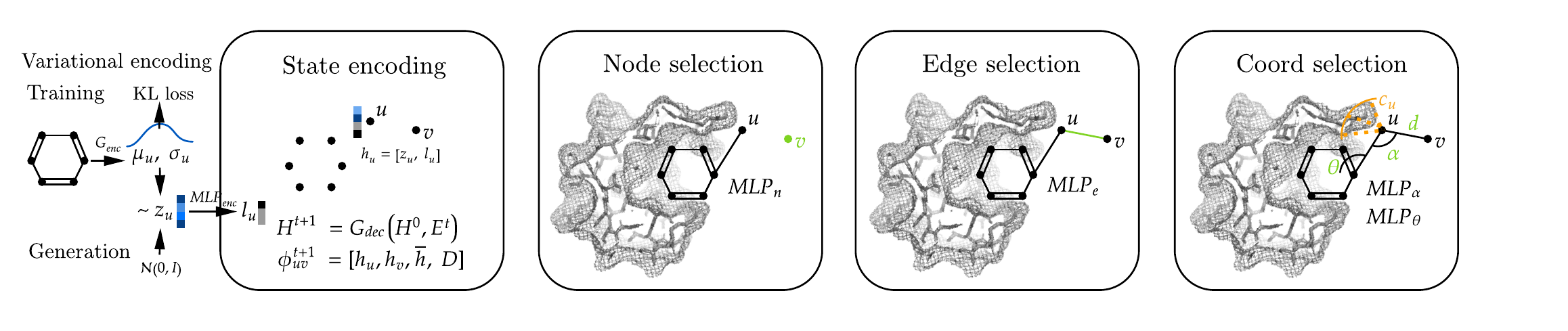}
\caption{Sequential autoencoder. The current partial molecule is encoded at each step, and next atom, bond, and node coordinate are selected based the molecular graph and structural pocket information.}
\label{fig:generation}
\end{figure}
\vspace*{-0.5cm}
\label{subsection:generation}
\subsection{Generation}
%The model is initialised with a number $N$ that determines the maximal number of atoms in the generated molecule. A latent embedding of the molecule, $\mathbf{Z}$, is obtained by sampling an encoding $\mathbf{z}_u \sim \mathcal{N}(0, I)$ for each of the $N$ atoms $u \in \mathbf{V}$. Starting from $\mathbf{Z}$, an entire molecule will be generated by the decoder, atom-by-atom, as follows.

Generation is seeded by sampling a latent encoding $\mathbf{z} \sim \mathcal{N}(\mathbf{0},\mathbf{I})$ for each of the atoms $u \in \mathbf{V}$ to give $\mathbf{Z} = \{\bm{z}_u\ | u \in \mathbf{V}\}$, where $|\mathbf{V}|:=N$ is a hyperparameter limiting the size of generated molecules. A molecule is generated atom-by-atom from $\mathbf{Z}$ as follows.

First, we decode an atom label $\bm{l}_u \in \mathbb{L}$ for each atom using a multi-layer-perceptron $\textup{MLP}_l$, which learns to map $\mathbf{Z}$ to labels $\mathbf{L}=\{\bm{l}_u | u \in V\}$. The latent sample $\bm{z}_u$ and label $\bm{l}_u$ together denote the features of the atoms $\bm{x}_u=[\bm{l}_u, \bm{z}_u]$. A graph-level encoding of the original graph, $\bar{\bm{x}}$, is also calculated by averaging the node latents. Next, we decode the molecule iteratively as a series of partial graphs $G^t(V^t,E^t)$, starting with $G^0(V, \emptyset)$. At each iteration $t$, the following steps are performed:

\begin{enumerate}[leftmargin=*]
    \item \textbf{State encoding} --- the current graph $G^t(V^t, E^t)$ is encoded using a gated graph neural network $\textup{GGNN}_{h}$. We produce node representations $\mathbf{H}^t = \{\bm{h}^t_u | u \in \mathbf{V}\}$ from the features $\mathbf{X}$ and current edges $E^t$ using iterated message passing with $\mathbf{H}_0^t = \mathbf{X}$ and $\mathbf{H}_{i+1}^t = \textup{GGNN}_h(\mathbf{H}_{i}^t, E^t)$. As in the CGVAE \cite{cgvae}, we calculate an average over all node latents, $\bar{\bm{h}}^t$, to represent the graph.
    \item \textbf{Node selection} --- To add a bond between a pair of atoms, we select source and target nodes. The source node $u$ is received from a queue in node construction order. The target node $v$ is chosen based on the state encoding and source node. For each possible target, we produce the feature vector $\bm{\phi}_{uv}^t=[\bm{h}_u^t, \bm{l}_u, \bm{h}_v^t, \bm{l}_v, \bar{\bm{x}}, \bar{\bm{h}}^t, d_{uv}, \mathbf{D}_{uv}]$ where $d_{uv}$ is the source-target graph distance, and $\mathbf{D}_{uv}$ represents novel coordinate information (see item \ref{item:coord_selection}). This vector encodes local and global structural information of the current and original molecules. A learned set operation maps $\mathbf{\Phi}^t = \{\bm{\phi}_{uv}^t | v \in V\}$ to a categorical distribution over nodes, and the maximum likelihood target is selected. If $v$ is the stop node, then we stop adding nodes to $u$ by removing it from the queue.
    \item \textbf{Edge selection} --- Having determined the edge $(u,v)$, we pick the bond type. We use another neural transformation $\textup{MLP}_e$ to map the edge feature vector $\bm{\phi}_{uv}^t$ to a distribution over bonds and select the label $\bm{e}_{uv}$ by maximum likelihood with masking to only select nodes $v$ that have free valency and to ensure bonds are only formed according to remaining valencies.
    \item \label{item:coord_selection} \textbf{Coordinate selection} --- To determine the coordinates of the newly added atom $v$, we enhance the state encoding with 3D structural information. We embed all atoms (pocket and ligand) into a latent SchNet \cite{schnet} representation $\bm{c}_u$ that captures the neighbourhood of the atoms in a rotation-and-translation invariant way useful for predicting chemical properties \cite{molgym}. We also add the selected edge type $\bm{e}_{uv}$ and obtain the feature vector $\bm{\psi}_{uv}^t = [\bm{e}_{uv}, \bm{h}_u^t, \bm{l}_u, \bm{h}_v^t, \bm{l}_v, \bm{c}_u^t, \bm{c}_v^t, \bar{\bm{c}}^t]$. We train $\textup{MLP}_{\alpha}$ and  $\textup{MLP}_{\theta}$ to predict angles and dihedral angles. The bond distance is given deterministically as a function of atoms and bond types $\bm{l}_u, \bm{l}_v, \bm{e}_{uv}$.
\end{enumerate}

\label{subsection:training}
\subsection{Training}
In training, the model is formulated as a variational autoencoder \cite{variational_bayes,cgvae}. We train on pairs of labeled ligand and pocket graphs ($G_l(V_l, E_l), G_p(V_p, E_p)$) from either the PDB or DUD-E datasets \cite{pdb_bind, dude}.

\paragraph{Encoding.} To match the intended use in generation, molecules are encoded as bags-of-nodes with features $\bm{z}_u$ sampled from $\mathcal{N}(\bm{\mu}_u, \textup{diag}(\bm{\sigma}_u^2))$ where the mean and variance are produced by a GGNN acting on $G_l$. We assume a standard normal prior on $\bm{z}$ resulting in the divergence term $\mathbb{KL}(\bm{z}_u || \mathcal{N}(\mathbf{0},\mathbf{I}))$ in the loss, which ensures that samples from $\mathcal{N}(\mathbf{0},\mathbf{I})$ are meaningful in generation.

% We encode a molecule by sampling the embedding of each atom $\mathbf{z}_u$ from a normal distribution $\mathcal{N}(\mu_u, \textup{diag}(\sigma_u^2))$. The mean and variance of the distribution are obtained using an encoder gated graph neural network $GGNN_{enc}$ on $G_l$. The encoder is constrained to produce parameters following the standard normal distribution using a KL loss term $L_{\text{KL}} = \sum_{u \in G_l} \mathbb{K}\mathbb{L}(\mathcal{N}(0,I) || \mathcal{N}(\mu_u, \textup{diag}(\sigma_u^2))$. Minimising the KL divergence term ensures that the atom embeddings sampled from $\mathcal{N}(0,I)$ during generation will remain meaningful to the decoder.

\paragraph{Decoding.}
Decoding is similar to generation, with added supervision at each step. We train using teacher forcing, forcing the decoder through a ground truth molecule construction sequence and minimising the negative log probability that the model assigns to the constructed molecule. A ground truth sequence consists of atom labels $\bm{l}^{*}_u$, and a sequence of partial graphs $G^{*0}, G^{*1}, \dots, G^{*T}$. We choose breadth-first construction sequences for training as they are natural for molecules \cite{cgvae}; moreover, they are equivalent to the queue-based sequential generation process and allow generation and training to share implementation logic.

First, we decode $\mathbf{Z}$ into node labels $\mathbf{L}$. This is supervised by minimising the log probability of re-generating the ground truth labels $\bm{l^*}$: $L_l = \sum_{u \in V_l} - \textup{log }  p(\bm{l}_u = \bm{l}^*_u | \mathbf{z})$. The other steps, which are either node, edge, or angle selections, are supervised in a similar fashion using the reconstruction loss $L_r = \sum_{t} - \textup{log }  p(G^t = G^{*t} |G^{*t-1})$. The overall minimised loss function is a weighted sum of individual terms $L = \lambda_r L_r + \lambda_l L_l + \lambda_{KL} L_{KL}$.

\label{section:results}
\section{Results}
\vspace*{-0.2cm}
We evaluate the generative model using a realistic drug design pipeline (Figure \ref{fig:evaluation_flow}, Appendix \ref{app:appendix_eval_pipeline}). The generated molecules are filtered against drug-likeness (QED) and synthesisability (SAS) scores, their conformers are optimised using RDKit distance geometry \cite{distance_geometry}, and docked against protein targets.

\label{subsection:qualitative_results}
\paragraph{Qualitative Results.}
We train our models on 4000 PDB pocket-ligand pairs with optimal hyperparameters available in Appendix \ref{app:appendix_hyperparams}. Example generated molecules and training evolution metrics are depicted in Appendix \ref{app:appendix_mol_gen}. The model is capable of reproducing similar ring structures and functional groups as in original encoded ligands, replicating CGVAE \cite{cgvae}. The similarity (Tanimoto) of generated molecules to original ligands increases over training, which shows that the model learns to encode and decode specific molecular features. We filter ill-formed molecules by discarding samples with drug-likeness (QED) $<0.20$, and synthesisability (SAS) $<8.0$. This results in receiving roughly the better 50\% of molecules with little runtime overheads. The filter is aligned with the SMINA docking benchmark \cite{dock} to make results more realistic, avoiding ill-formed molecules that take long to dock or cause docking failures.

\label{subsection:gen_benchmark}
\paragraph{Coordinate Generation Benchmark.}
Besides molecular graphs, we also evaluate the quality of generated coordinates. MolGym (\cite{molgym}) provides an environment for benchmarking conformers based on quantum-physical energy calculations. The environment rewards each atom placement step $t$ with the amount of energy that was released when forming the bond $-\Delta E = E^{t} - (E^{t-1} + E_{\text{new\_atom}})$. The authors of MolGym directly maximise the reward using reinforcement learning, which we use as a comparison baseline. MolGym defines two tasks, single and multi-bag. The coordinate quality of our generative model improves with training on all tasks as shown in Figure \ref{fig:training_metrics_molgym}, Appendix \ref{app:appendix_coord_bench}. Some features of generated 3D structures are highlighted in Figure \ref{fig:3d_structures}, Appendix \ref{app:appendix_coord_bench}. The model is successful in learning to generate coordinates and implicitly improves the energy reward, as a side effect of learning to replicate molecules. Due to the discretised coordinate space, the 3D conformers are necessarily approximate and often contain errors that accumulate as more atoms are placed. However, the aim is not to generate valid conformers, but provide a differentiable mechanism to guide molecule generation with an approximate location of the atoms within the pocket. We refine the conformers afterwards (before docking) using RDKit's distance geometry methods \cite{distance_geometry}.

\label{subsection:dock_benchmark}
\paragraph{Docking Benchmark.}
We compare the performance of guiding generation with coordinates to a baseline that only generates 2D graphs without pocket information. We measure the Vinardo \cite{vinardo} docking scores of generated molecules, which was proposed as a realistic benchmark for structure-based design (SMINA benchmark \cite{dock}). We aggregate drug-likeness (QED), synthetic accessibility (SAS), and docking (DOCK) scores of molecules in the following design task --- generate/design a ligand based on a randomly chosen target protein pocket. The experiment is performed 200 times and docking scores of designed ligands in target pockets are recorded. The full docking benchmark with additional design tasks is summarised in Appendix \ref{app:appendix_dock_bench}.

\begin{minipage}{\textwidth}
    \begin{minipage}[b]{0.49\textwidth}
    \centering
    \begin{tabular}{@{}lccccccccl@{}}
    \toprule
    \multirow{2}{*}{Model} & \multicolumn{3}{c}{Multi-mol design}
    \\\cmidrule(lr){2-4} & DOCK $\downarrow$ & QED $\uparrow$ & SAS $\downarrow$ \\
    \midrule
    Structure Guided       & \textbf{-4.41} & \textbf{0.36} & 5.81\\
    Unguided     & -4.15 & 0.32 & \textbf{5.57}\\
    \midrule
    PDB Average  & -5.32 & 0.41 & 4.58\\
    Orig. Ligand & -8.87 & 0.41 & 4.58\\
    \bottomrule
    \end{tabular}
    \captionof{table}{Coordinate-guided versus unguided generation results compared to PDB statistics.}
    \label{tab:guided_unguided}
    \end{minipage}
    \hfill
    \begin{minipage}[b]{0.49\textwidth}
    \centering
    \includegraphics[scale=0.12,trim={50cm 0cm 0cm 55cm},clip]{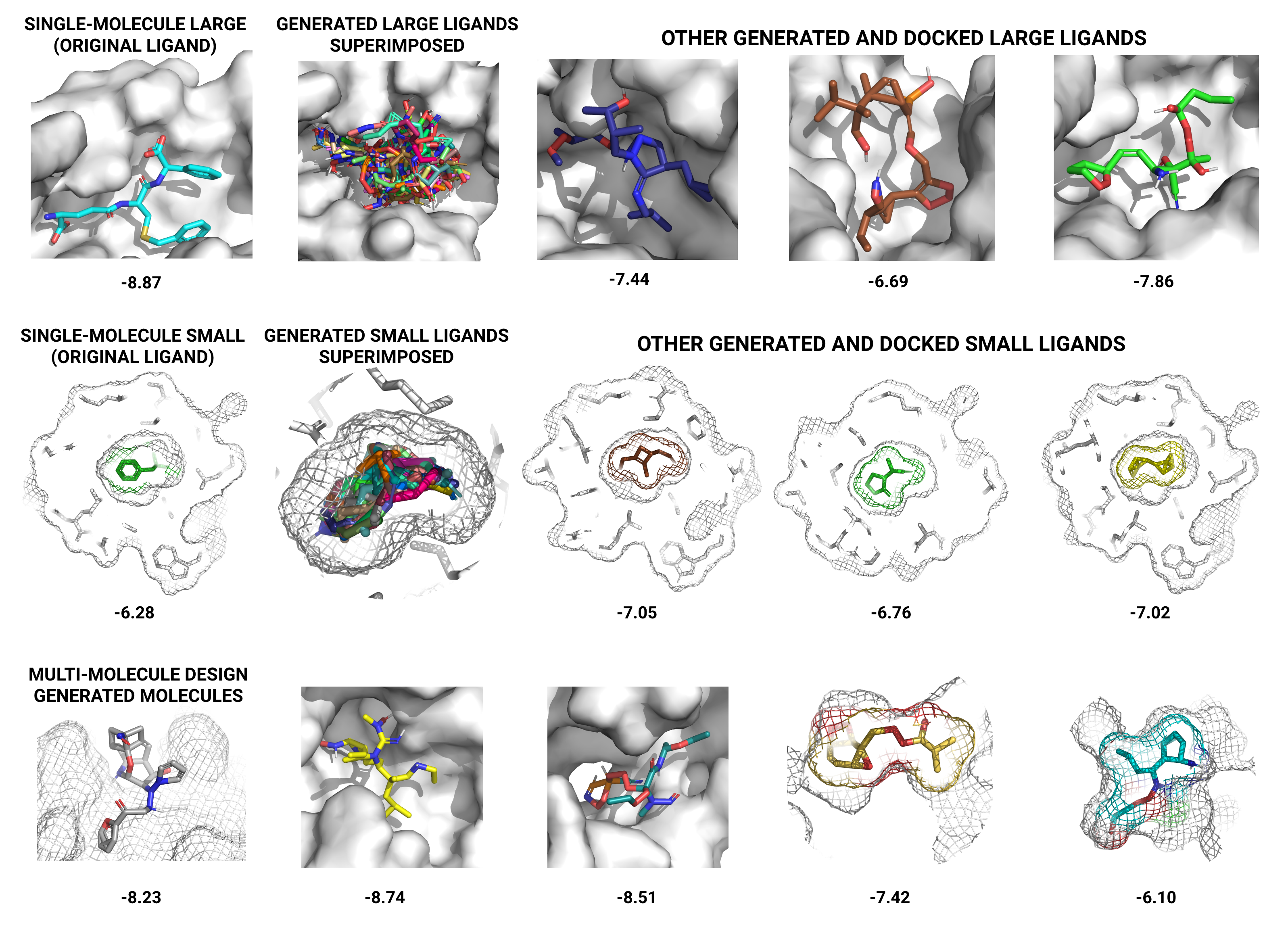}
    \captionof{figure}{Generated drug candidates with high predicted binding affinities (top 10\%).}
    \label{fig:docked_ligands}
  \end{minipage}
\end{minipage}

The guided generative method outperforms the unguided model by $6\%$ on docking scores and $10\%$ in the average multi-molecule design task (Table \ref{tab:guided_unguided}). We observed that the guided generator produces more rings, which is known to make more rigid and stable-bound molecular conformations. On the other hand, large rings are harder to synthesise, which accounts for the harder average synthesisability of molecules produced during guided generation; however, these molecules can be cheaply filtered. Overall, there are generated molecules that dock better than the original PDB ligands, which could be submitted for wet-lab testing for functional validation.

\label{section:conclusion}
\section{Conclusion}
\vspace*{-0.2cm}
We presented a novel approach to ligand generation that is aware of protein pocket structure and achieves higher docking scores. However, 3D structure is merely one aspect of the complex binding problem and we believe that more ligand-pocket interactions and dynamics should be considered. Future work may enhance the model's pocket encodings to capture interactions such as hydrogen bonds, repulsion, and hydrophobicity, as additional atom features.

% from thesis
% As suggested by the evaluation findings, there is potential % to incorporate chemical 
% complementarity and protein-ligand %interactions, besides structural properties. Docking 
% scores such as Vinardo [32] use hydrogen bonds and electrostatic repulsion to better model 
% realistic binding scores. Such information could be added into my generative model as features of atoms

\printbibliography

\appendix
\label{app:appendix_hyperparams}
\section{Hyperparameters}
\begin{table}[h]
    \centering
    \begin{tabular}{ll}
        \toprule
        Hyperparameter & Values \\
        \midrule
        Batch size & 1, 4, 8, \textbf{16}, 32\\
        Epochs & 1-7, \textbf{8-10}, 10-20\\
        Encoder & Non-Variational, \textbf{Variational}\\
        Graph NN & GAT, \textbf{GGNN}\\
        Propagation steps & 1-6, \textbf{7}, 8-9 \\
        Embedding size & 30, 50,  \textbf{80}, 100, 150\\
        SchNet Embedding size & \textbf{16}, 32, 64\\
        \bottomrule
    \end{tabular}
    \vspace{0.1in}
    \caption{Evaluated hyperparameters with best performance highlighted in \textbf{bold}.}
    \label{tab:hyperparams}
\end{table}

\label{app:appendix_coords}
\section{Internal coordinate system}
\begin{figure}[h]
\includegraphics[scale=1]{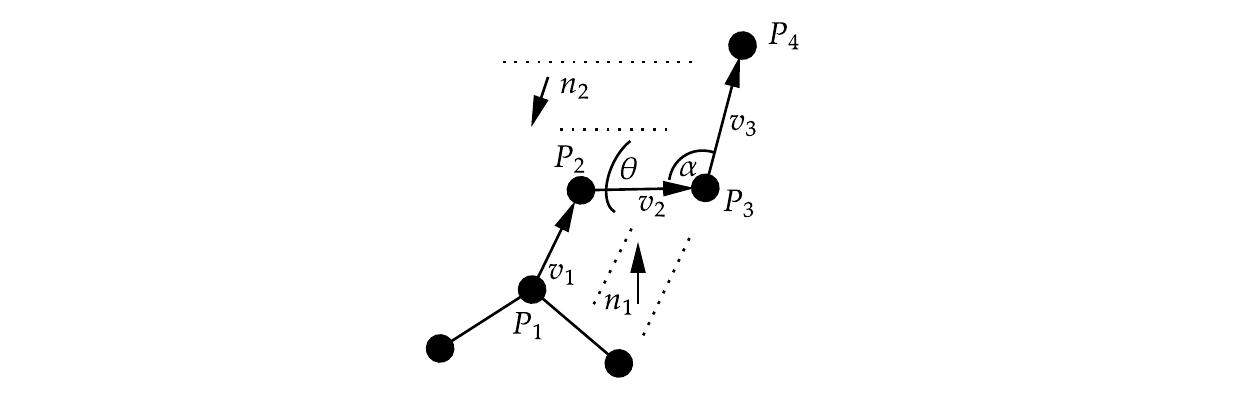}
\caption{Internal coordinate representation of a molecule. Atom at position $P_4$ is represented relative to the closest atoms $P_3, P_2, P_1$.}
\label{fig:torsion}
\end{figure}
The internal coordinate system is shown in Figure \ref{fig:torsion}. Consider describing the position of the atom located at $P_4$ using the three closest atoms at locations $P_3, P_2, P_1$ (in that order), and let $P_{i+1}-P_{i}=v_i$. The point $P_4$ is uniquely described using its distance $|v_3|$ from the nearest atom at $P_3$, the angle $\alpha$ between vectors $v_3v_2$, and dihedral angle $\theta$ between the planes with normals $n_1 = \widehat{v_2 \times v_1}$ and $n_2 = \widehat{v_3 \times v_2}$ ($\hat{v} = v/|v|$ denotes normalisation). A molecule can be converted from cartesian to internal coordinates by traversing the atoms in a breadth-first manner and calculating $d, \alpha, \theta$ using the following trigonometric equations:
\begin{equation}
\alpha = \textup{atan2}(\widehat{v2 \times v3}, v_2 \cdot v_3)
\end{equation}
\begin{equation}
\theta = \textup{atan2}(\widehat{n_1 \times \hat{v}_2} \cdot n_2, n_1 \cdot n_2)
\end{equation}
\begin{equation}
d = |v3|
\end{equation}
For conversion back from internal to absolute cartesian coordinates, one needs to choose a starting point and orientation of the molecule. By convention, the first atom $P_0$ is placed at the origin $O$, the second atom $P_1$ is placed bond distance $d_{O_0P_1}$ away from the first atom in the $z$ direction. The third atom is placed in the $xz$ plane, distance $d_{P_1P_2}$ from $P_1$, with angle $\alpha$ between lines $P_0P_1$ and $P_1P_2$. The fourth and subsequent atom coordinates are determined by the equations:
\begin{equation}
v_3 = \textup{cos}(\alpha)\hat{v}_2 + \textup{sin}(\alpha)\textup{cos}(\theta)(n_1\times \hat{v}_2)
\end{equation}
\begin{equation}
P_4 = P_3 + d\hat{v}_3
\end{equation}

\label{app:appendix_mol_gen}
\section{Molecule generation}
\begin{figure}[htp]
\centering
\hspace*{-0.3cm}
\includegraphics[scale=0.39]{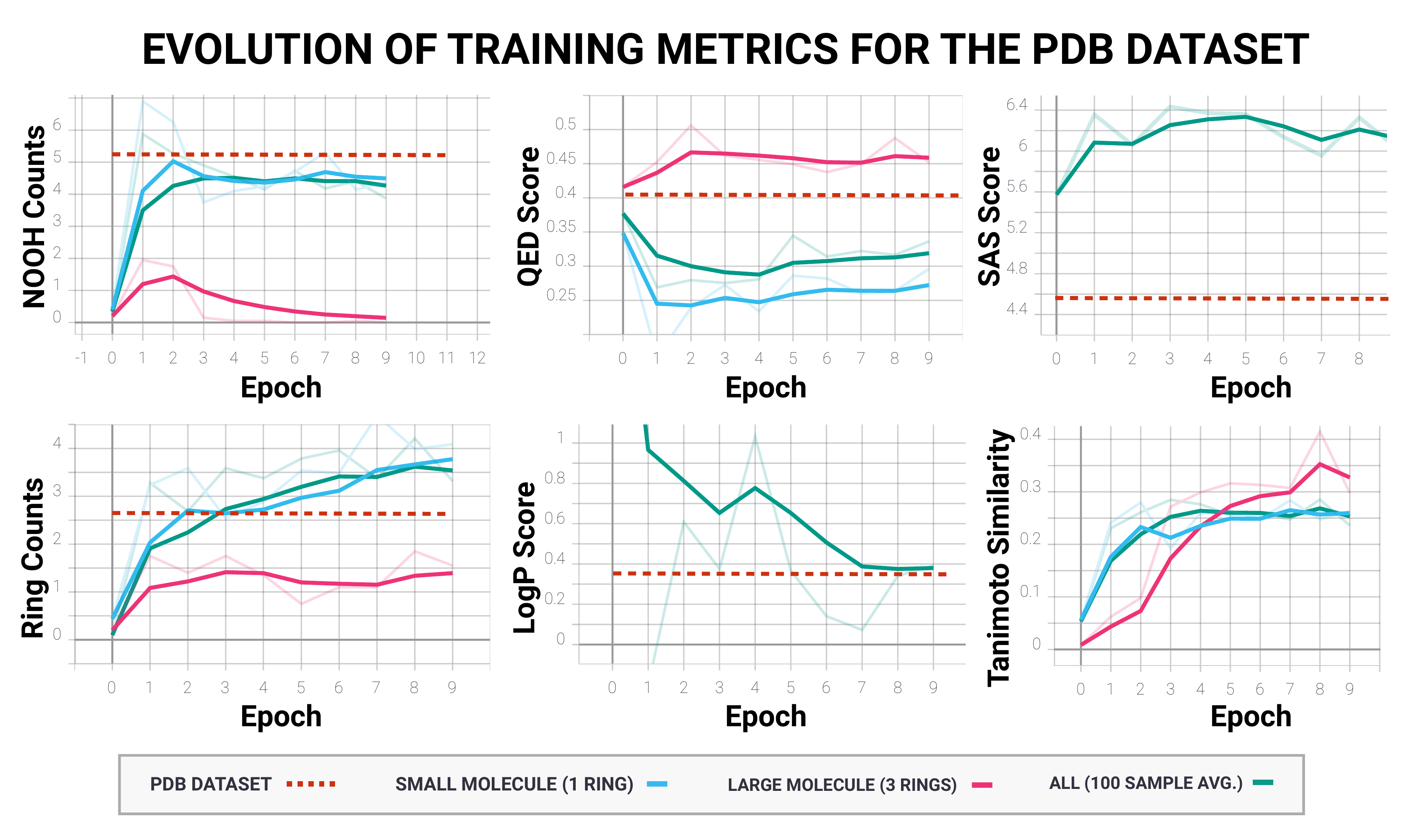}
\caption{Evolution of molecular metrics --- drug-likeness (QED), synthesisability (SAS), similarity (Tanimoto), and partition coefficient (LogP), and functional group counts (rings and NOON).}
\label{fig:training}
\end{figure}

The model is capable of reproducing similar ring structures and functional groups present in original encoded ligands and replicates the variational behaviour of the CGVAE autoencoder \cite{cgvae}. Figure \ref{fig:training} shows the evolution of model metrics during training. The counts of "NO" and "OH" functional groups match the dataset. Ring counts are optimal for both the small and large molecule and overshoot the dataset statistics in average. This is due to ill-formed molecules with large rings, which are filtered later.

The similarity (Tanimoto) of generated molecules to original ligands increases over training to $25\%$, which shows that the model learns to encode and decode specific molecular features. The LogP score is always in the Lipinski druglikeness range \cite{lipinski}, but usually has a high variance. Although valency masking ensures syntactical correctness of molecules, there are generated samples which cannot be syntehesised due to large rings, or long bonds.

\begin{figure}[htp]
\centering
\hspace*{-0.8cm}
\includegraphics[scale=0.15]{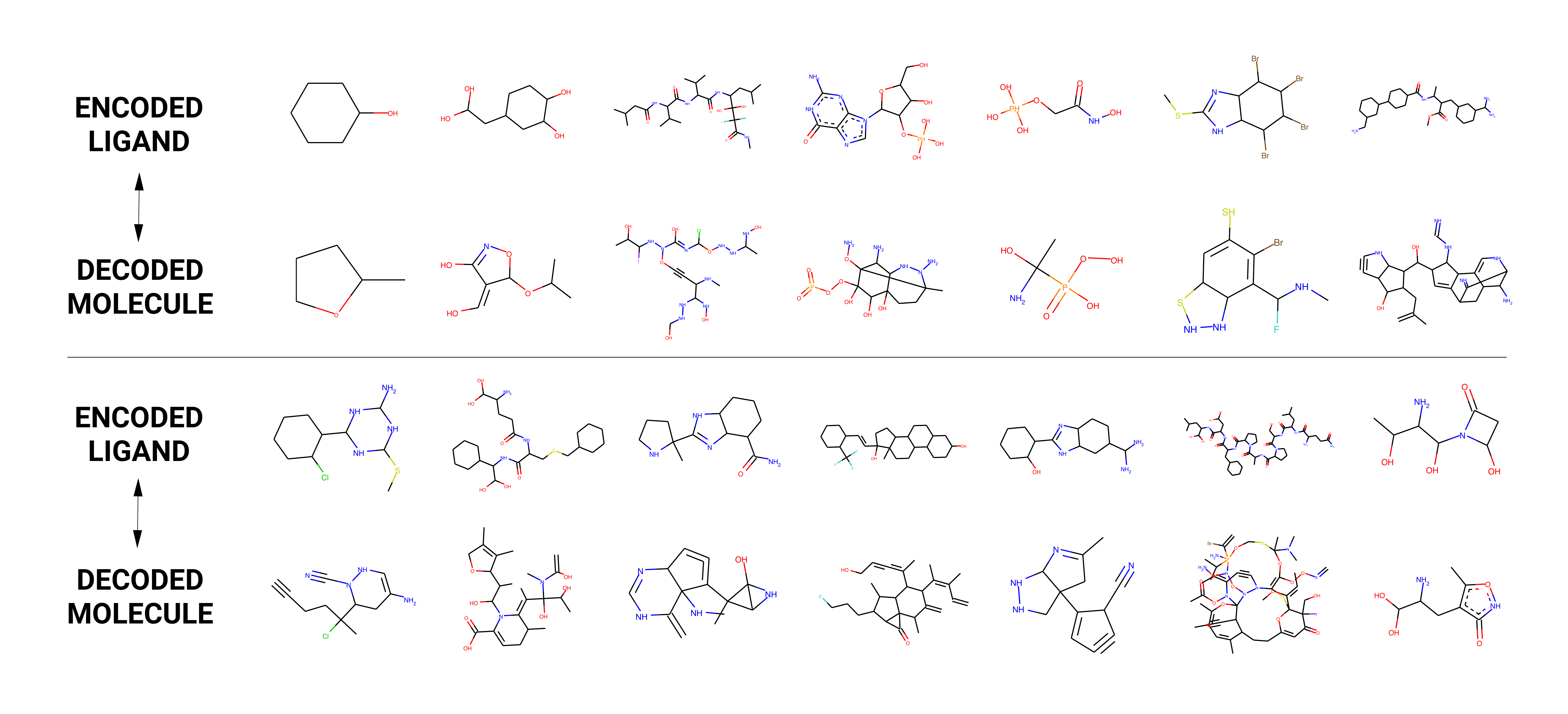}
\caption{Examples of predicted molecules with similar qualities to original ligands.}
\label{fig:predictions}
\end{figure}

Example molecules sampled by the model trained for 8 epochs are highlighted in Figure \ref{fig:predictions}. Some molecules show remarkable similarity to original ligands. During early training (epochs 1--3), the model generates a high proportion of incorrect molecules such as in Figure \ref{fig:defects}, and starts capturing meaningful similarity towards epochs 6--7. Some erroneous molecules are always produced, mainly when the encoded ligand contains a large number of atoms and rings.

We observed that commonly used metrics such as similarity often do not correlate with the quality of molecules. A corrupt molecule that contains numerous functional groups (Figure \ref{fig:qualitative_errors} left) can be ranked similarly to the original ligand and large numbers of rings can cause false high QED scores. Hence we always use multiple metrics in evaluation. Some low scores are caused by defects in molecules that naturally occur in variational approaches \cite{delinker} and would likely be un-learnt by pretraining on a large dataset such as ZINC \cite{zinc} used in CGVAE \cite{cgvae}. 

\label{app:appendix_eval_pipeline}
\section{Evaluation Pipeline}
\vspace*{1cm}
\begin{figure}[htp]
\centering
\includegraphics[scale=0.32]{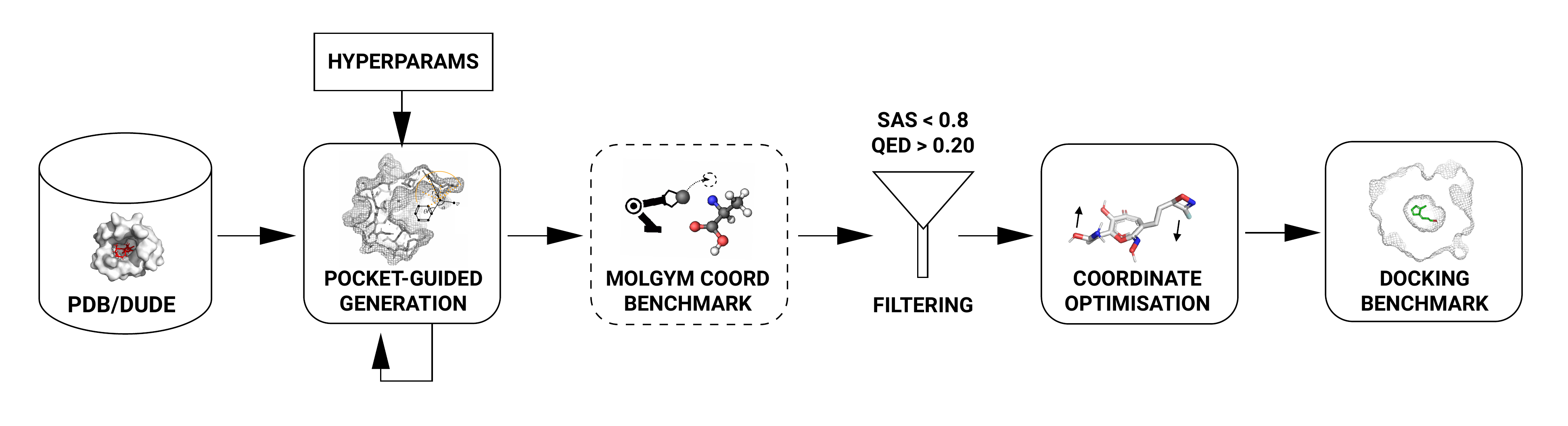}
\caption{Evaluation pipeline.}
\label{fig:evaluation_flow}
\end{figure}

\label{app:appendix_coord_bench}
\section{Coordinate Benchmark}
\begin{figure}[htp]
\centering
\hspace*{-0.2in}
\includegraphics[scale=0.32]{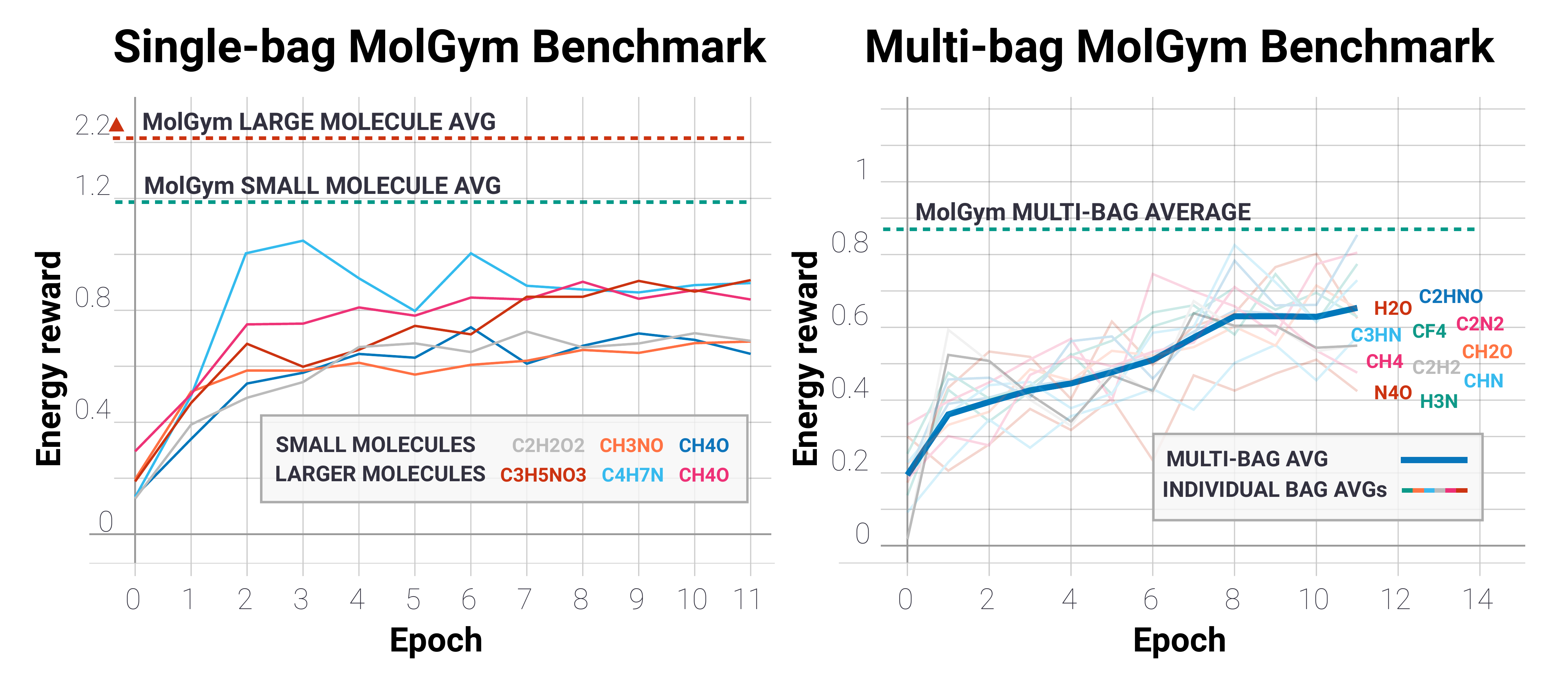}
\vspace*{-0.5cm}
\caption{Evolution of molecular energy reward with training.}
\label{fig:training_metrics_molgym}
\end{figure}

We train our model on 1000 PDB protein-ligand pairs and evaluate its performance on the two tasks defined by MolGym.
\begin{itemize}[leftmargin=*]
    \item \textbf{Single-bag task} --- the task is to construct a molecule out of a bag of atoms such as $\{C: 1, H: 4, O: 1\}$. Three small bags and three larger bags are available and we generate $30$ molecules and plot average rewards for each bag.
    \item \textbf{Multi-bag task} --- given a multi-bag of 11 bags of atoms, the task is to construct stable molecules with an arbitrary bag chosen for each run. We generate $11 \times 30$ molecules after each epoch (30 on average for each bag), and plot the overall average reward and rewards for individual bags.
\end{itemize}
The evolution of rewards over training epochs is plotted in Figure \ref{fig:training_metrics_molgym}. The model generates a range of structures that resemble existing molecules. The best performance is observed when generating coordinates of longer carbon chains without rings, or generating smaller molecules. The model often does not close ring structures correctly, which leads to accumulating errors in predicted coordinates afterwards. This loop closure problem is well studied in other fields, such as localisation. A potential solution could be using batch normalisation or a backtracking algorithm to periodically correct for errors after a few generative steps, as it is impossible to foresee whether the currently generated atom will form a ring.

Note that direct comparisons to MolGym \cite{molgym} are unfair in both tasks due to the following differences: 1. MolGym is trained to directly maximise the provided reward, while our model is trained to reproduce construction steps of existing molecules, 2. MolGym is trained separately for each provided bag or multi-bag, while our model is always trained on a 1000 existing ligands and has not seen the specific testing bags, 3. our method discretises molecular space, which can result in small atom misplacements impacting binding energies, where in reality the molecule would flexibly adjust.

\begin{figure}[h]
\centering
\includegraphics[scale=0.125]{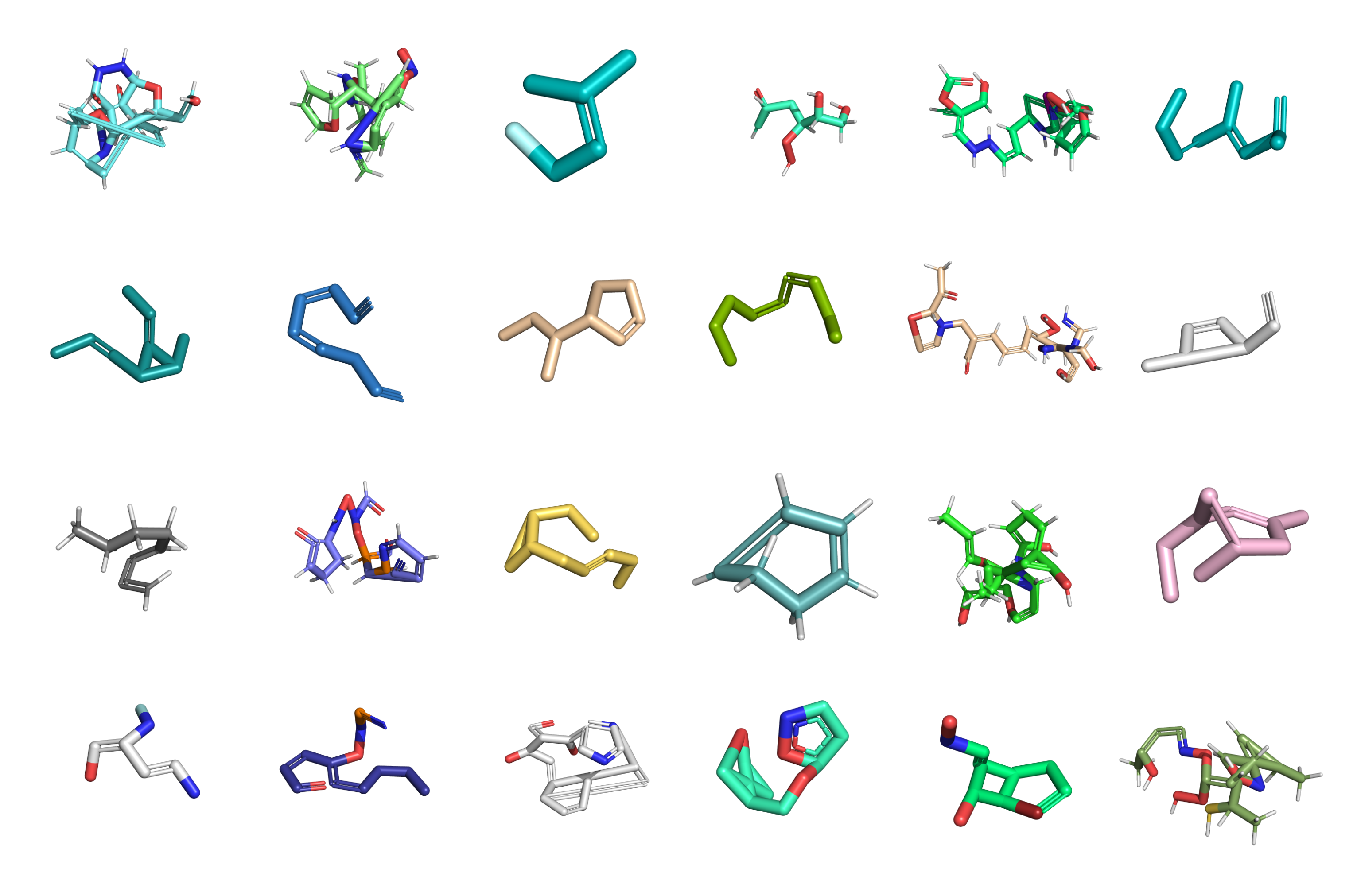}
\caption{Illustration of 3D coordinates generated by the model, highlighting some common issues caused by rings. These structures are used to guide generation and optimised before docking. (Colour scheme: C - dominant, N - blue, O - red)}
\label{fig:3d_structures}
\end{figure}

\label{app:appendix_dock_bench}
\section{Docking Benchmark Full}
We aggregate drug-likeness (QED), synthetic accessibility (SAS), and docking (DOCK) scores of molecules in three design tasks resembling the docking benchmark:
\begin{itemize}[leftmargin=*]
    \item \textbf{Multi-molecule design} --- the task is to generate/design a ligand based on a randomly chosen encoded protein-ligand pair. The experiment is performed 200 times and docking scores of designed ligands and target are recorded.
    \item \textbf{Single-molecule large ligand design} --- 100 ligands generated based on a single protein-ligand pair with docking scores measured for the single pocket.
    \item \textbf{Single-molecule small ligand design} --- same as above, with a small protein.
\end{itemize}

\begin{table}[h]
    \small
    \centering
    \begin{tabular}{@{}lccccccccl@{}}
    \toprule
    \multirow{2}{*}{Model} & \multicolumn{3}{c}{Multi-mol design} & \multicolumn{3}{c}{Single-mol design large} & \multicolumn{3}{c}{Single-mol design small}
    \\\cmidrule(lr){2-4}\cmidrule(lr){5-7}\cmidrule(lr){8-10}
                        & DOCK& QED& SAS       & DOCK      & QED       & SAS       & DOCK      & QED       & SAS    \\
    \midrule
    Guided       & \textbf{-4.41} & \textbf{0.36} & 5.81 & -2.81 & 0.24 & 6.42 & -4.72 & \textbf{0.42} & \textbf{5.17}\\
    Unguided     & -4.15 & 0.32 & \textbf{5.57} & \textbf{-3.5} & \textbf{0.26} & \textbf{6.06} & -4.72 & 0.41 & 5.19\\
    \midrule
    PDB Average  & -5.32 & 0.41 & 4.58 & -6.06 & 0.18 & 4.03 & -3.7 & 0.55 & 1.83\\
    Orig. Ligand & -8.87 & 0.41 & 4.58 & -10.12 & 0.18 & 4.03 & -6.28 & 0.55 & 1.83\\
    \bottomrule
    \end{tabular}
    \vspace{0.1in}
    \caption{Coordinate-guided versus unguided generation results compared to original PDB protein-ligand pair scores, and average PDB scores. Docking scores ($\textup{DOCK}$, $\textup{kcal}/\textup{mol}$) and synthesisability scores ($\textup{SAS} \in [1,10]$) should be low, drug-likeness scores ($\textup{QED} \in [0,1]$ should be high. PDB Average and Original Ligand scores are identical when the score is pocket-independent (QED, SAS).}
    \label{tab:guided_unguided}
\end{table}

\begin{figure}[h]
\centering
\includegraphics[scale=0.12]{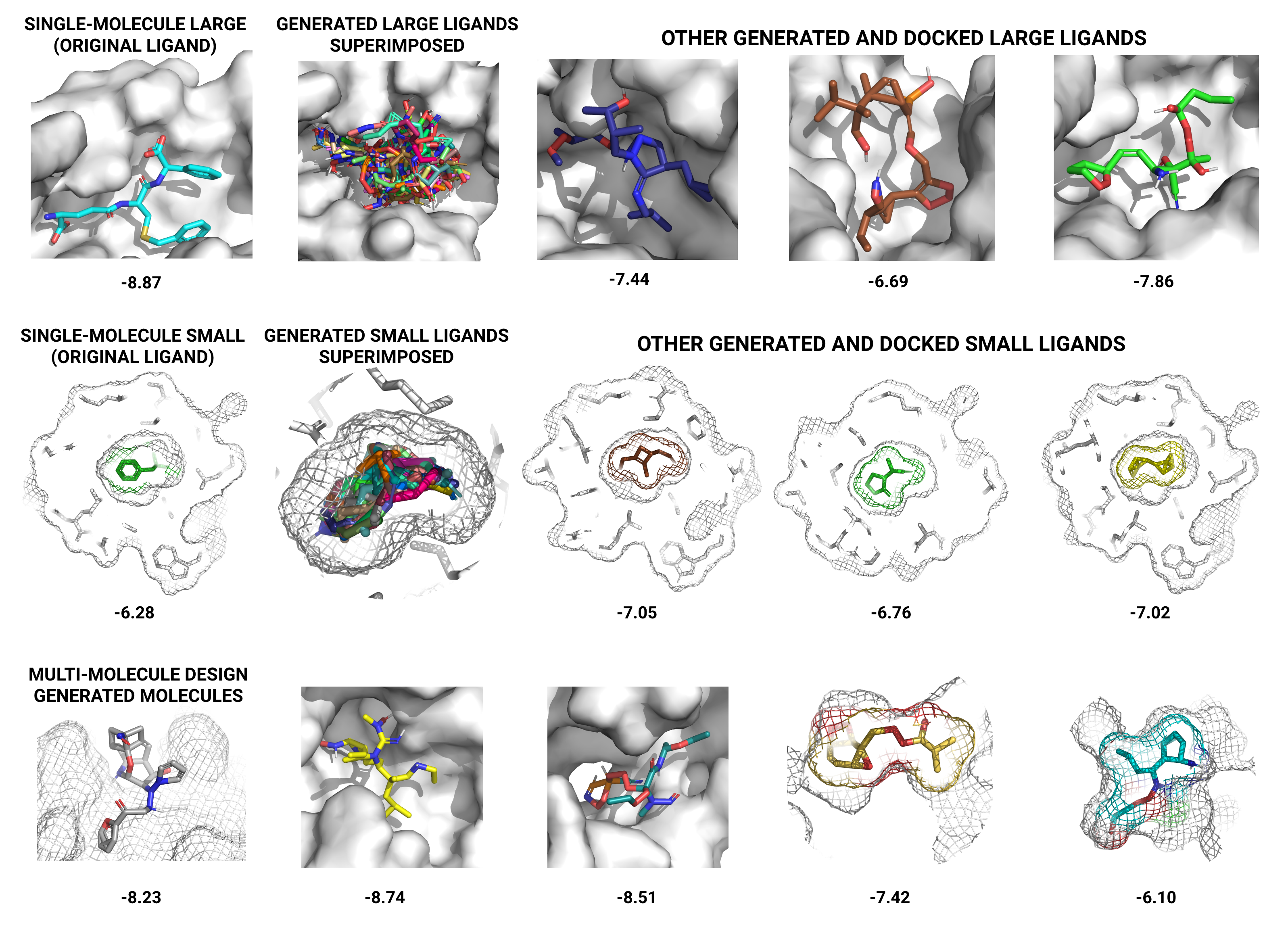}
\caption{Generated drug candidates with high binding affinities (top 10\%).}
\label{fig:docked_ligands}
\end{figure}

\begin{figure}[htp]
\centering
\hspace*{-0.6cm}
\includegraphics[scale=0.12]{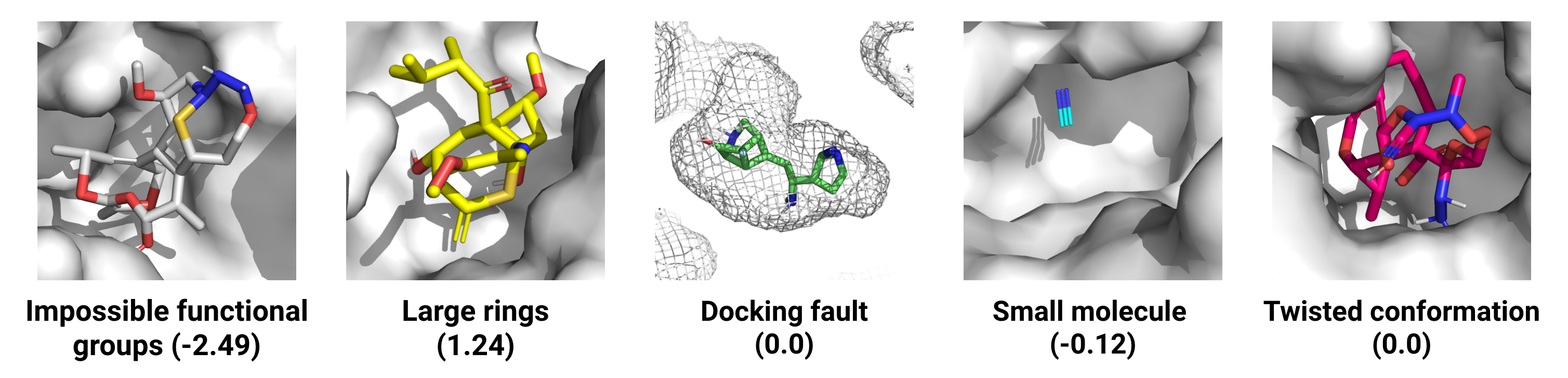}
\caption{Commonly generated defects with low binding affinities (bottom 20\%).}
\label{fig:defects}
\end{figure}
The guided generative method outperforms the unguided model by $6\%$ on docking scores and $10\%$ in the average multi-molecule design task (Table \ref{tab:guided_unguided}). The top $10\%$ of ligands generated to resemble the small ligand achieved lower (better) binding affinities than the original ligand in PDB, while coming close to the original for larger ligands as well. Example structures docked to proteins after coordinate optimisation are shown for the three design tasks in Figure \ref{fig:docked_ligands}, with their docking scores displayed below. Both models are capable of generating molecules that dock better or close to the original ligands, demonstrating the potential usefulness of the framework in realistic drug design. The superimposed figures demonstrate that the models explore many ways of fitting the same pocket.

Example structures that do not bind well (bottom 20\% of generated molecules) are shown in Figure \ref{fig:defects}. Common cases for high binding affinities seem to be small molecules and large rings, which can be effectively filtered in the future. Another common issue are twisted structures, which are penalised by the Vinardo algorithm as they do not fill space optimally, potentially causing docking to fail completely.

\end{document}